\documentclass[a4paper,superscriptaddress]{jpconf}

\usepackage{graphicx}
\usepackage{amsmath}
\usepackage{amssymb}
\usepackage{mathtext}

\begin{document}
\title{Specific interface area in a thin layer system of\\ two immiscible liquids with vapour generation at\\ the contact interface}

\author{Anastasiya V Pimenova$^1$, Ilias M Gazdaliev$^2$ and Denis S Goldobin$^{1,2}$}
\address{$^1$Institute of Continuous Media Mechanics UB RAS,
         1 Akademika Koroleva street, 614013~Perm,~Russia}
\address{$^2$Department of Theoretical Physics, Perm State University,
         15 Bukireva street, 614990~Perm,~Russia}
\ead{Anastasiya.Pimenova@gmail.com, Denis.Goldobin@gmail.com}

\begin{abstract}
For well-stirred multiphase fluid systems the mean interface area per unit volume, {\it or} ``specific interface area'' $S_V$, is a significant characteristic of the system state. In particular, it is important for the dynamics of systems of immiscible liquids experiencing interfacial boiling. We estimate the value of parameter $S_V$ as a function of the heat influx $\dot{Q}_V$ to the system or the average system overheat $\langle\Theta\rangle$ above the interfacial boiling point. The derived results can be reformulated for the case of an endothermic chemical reaction between two liquid reagents with the gaseous form of one of the reaction products. The final results are restricted to the case of thin layers, where the potential gravitational energy of bubbles leaving the contact interface is small compared to their surface tension energy.
\end{abstract}

\section{Introduction}

For multiphase liquid systems the mean contact interface area per unit volume, {\it or} the ``specific interface area'' $S_V\equiv(\delta{S}/\delta{V})$, is an important characteristic of the state. It is especially significant for the systems where this interface is active chemically or in some other way. The systems of immiscible liquids experiencing interfacial boiling~\cite{Krell-1982,Geankoplis-2003,Simpson-etal-1974,Celata-1995,Roesle-Kulacki-2012-1,Roesle-Kulacki-2012-2,Sideman-Isenberg-1967,Kendoush-2004,Filipczak-etal-2011,Gordon-etal-1961,Prakash-Pinder-1967-1,Prakash-Pinder-1967-2,Pimenova-Goldobin-JETP-2014,Pimenova-Goldobin-EPJE-2014} are an example of the systems where parameter $S_V$ becomes especially important. For the interfacial boiling the vapour phase grows if the sum of saturated vapour pressures of the liquids brought in contact excesses the atmospheric pressure, which means that this boiling can occur under conditions where no boiling in the bulk of both liquids is possible.  The problem of calculation of $S_V$ cannot be addressed rigourously and any direct numerical simulation, being extremely challenging and CPU-time consuming, will provide results pertaining to specific system set-ups. Thus, some general assessments on $S_V$ can be highly beneficial.

In this paper we provide the assessments for the process of gas generation at the direct contact interface for a thin layer. The generated vapour volume in the system will be proportional to the heat influx to the interface. Therefore, our results will be relevant both for the case of endothermic chemical reaction between two liquids with the gaseous form of one of the reaction products and for the case of interfacial boiling~\cite{Pimenova-Goldobin-EPJE-2014,Goldobin-Pimenova-2017}. At the direct contact interface, a vapour layer grows and produces bubbles which breakaway of the interface and rise. The presence of vapour bubbles changes the fluid buoyancy and performs a ``stirring'' of the system. Additional kinetic energy is brought into the system by bubble poppings at the liquid--atmosphere interface. In~\cite{Goldobin-Pimenova-2017}, the theory has been developed for the case where one can neglect the latter contribution into the balance of mechanical energy. In this work we consider an opposite limiting case; the liquid layer is thin in a sense that the potential gravitational energy of rising bubbles is small compared to their surface tension energy.

\begin{figure}[!t]
\begin{minipage}[b]{3.5in}
\center{
\includegraphics[width=0.523\textwidth]%
{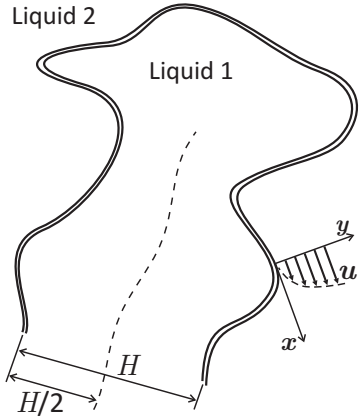}
}
\end{minipage}
\quad
\begin{minipage}[b]{2.4in}
 \caption{Sketch of the vapour layer between two immiscible liquids and the average liquid flow $u$ within the turbulent boundary layer
}
\label{fig1}
\vspace{10pt}
\end{minipage}
\end{figure}

For a two-liquid system experiencing direct contact boiling, the quantity of our interest depends on parameters of liquids and characteristics of the evaporation process, which are controlled by the mean overheating and the bubble production rate~\cite{Filipczak-etal-2011}. In this work the volumes of both components are assumed to be commensurable, no phase can be considered as a medium hosting dilute inclusions of the other phase. The characteristic width of the neighborhood of the vapour layer, beyond which the neighborhood of another vapour layer lies, is (see figure~\ref{fig1})
\begin{equation}
\nonumber
H\sim S_V^{-1}.
\end{equation}

The process of boiling of a mixture above the bulk boiling temperature of the more volatile liquid is well-addressed in the literature~\cite{Simpson-etal-1974,Celata-1995,Roesle-Kulacki-2012-1,Roesle-Kulacki-2012-2,Sideman-Isenberg-1967,Kendoush-2004,Filipczak-etal-2011}. Hydrodynamic aspects of the process of boiling below the bulk boiling temperature~\cite{Pimenova-Goldobin-JETP-2014,Pimenova-Goldobin-EPJE-2014} have to be essentially similar at the macroscopic level; rising and popping vapour bubbles drive the stirring of system, working against the gravitational stratification into two layers with a flat horizontal interface, the surface tension forces tending to minimize the interface area, and viscous dissipation of the flow kinetic energy. Specifically, the behaviour of parameter $S_V$ depending on macroscopic characteristics of processes in the system should be the same as for systems with superheating of the more volatile component. In what follows, we perform an analytical assessment of the dependence of $S_V$ on the evaporation rate for a well-stirred system. The evaporation rate can be controlled either by the average overheat $\Theta$ of the system above the interfacial boiling point $T_\ast$ (section~4) or heat influx (section~5).

\section{Energy flux balance in a well-stirred system}
Let us derive the rough relationships between the macroscopic parameter $S_V$ of the system state and the heat influx rate per unit volume $\dot{Q}_V=\delta{Q}/(\delta{V}\delta{t})$ for a statistically stationary process of interfacial boiling.

The flow and consequent stirring in the system are enforced by the buoyancy and poppings of the vapour bubbles, while other mechanisms counteract the stirring of the system. These other mechanisms are gravitational stratification of two liquids, surface tension tending to minimise the interface area and viscous dissipation of the flow energy. Since the latent heat of phase transitions and heat of temperature inhomogeneities are enormously large compared to the realistic values of the kinetic energy of microscopic motion and gravitational potential energy ({\it e.g.}, see~\cite{Goldobin-Pimenova-2017} for quantitative estimates), the latter can be neglected in consideration of the heat balance. Hence, all the heat inflow into the system can be considered to be spent for the vapour generation;
$\dot{Q}_VV\rightarrow(\Lambda_1n_1^{(0)}+\Lambda_2n_2^{(0)})\dot{V}_\mathrm{v}$,
where $V$ is the system volume, $\dot{V}_\mathrm{v}$ is the volume of the vapour produced in the system per unit time, $\Lambda_j$ is the enthalpy of vaporization per one molecule of liquid $j$, and $n_j^{(0)}$ is the saturated vapour pressure of liquid $j$. Thus,
\begin{align}
\dot{V}_\mathrm{v}=\frac{\dot{Q}_V\,V}
{\Lambda_1n_1^{(0)}+\Lambda_2n_2^{(0)}}\,.
\label{eq2-01}
\end{align}

The mechanical energy income from one rising and popping vapour bubble $E_\mathrm{bub}$ is converted into the kinetic energy of liquid flow, the potential energy of a stirred state of the two-liquid system, the surface tension energy and dissipated by viscosity forces. In a statistically stationary state, the mechanical kinetic and potential energies do not change averagely over time and all the energy influx is to be dissipated by viscosity;
$$
E_\mathrm{bub}N\longrightarrow\dot{W}_\mathrm{l,k}\tau\,,
$$
where $\dot{W}_\mathrm{l,k}$ is the rate of viscous dissipation of energy, $\tau$ is the time of generation of the vapour volume $V_\mathrm{v}$, $V_\mathrm{v}=\dot{V}_\mathrm{v}\tau$,  $N$ is the number of bubbles which reach the atmosphere--liquid interface per time $\tau$. Hence,
\begin{align}
E_\mathrm{bub}\frac{\dot{V}_\mathrm{v}}{V_\mathrm{bub}}\longrightarrow\dot{W}_\mathrm{l,k}\,,
\label{eq2-02}
\end{align}
where $V_\mathrm{bub}$ is the characteristic volume of a bubble.

\subsection{Energy income $E_\mathrm{bub}$ from one bubble}
The average potential gravitational energy of the rising bubble is $\rho g V_\mathrm{bub}h/2$, where we assume a nearly uniform distribution of the bubble formation sites over height, $h$ is the liquid layer thickness, $\rho$ is the characteristic density of two liquids.

The surface tension energy of the bubble is $4\pi r_\mathrm{bub}^2\sigma$, where $\sigma$ is the characteristic surface tension for two liquids. Here, one should also take into account the energy of the vapour compression by the surface tension forces, which will be released as the bubble pops. The pressure increase by the surface tension forces is $\delta P=2\sigma/r_\mathrm{bub}$. As the bubble pops, the pressure excess over the atmospheric pressure drops to zero, and the work made by the expanding vapour is $A_{\delta V}=\int\!P\,\mathrm{d}V$. For a reference bubble radius $\sim1\,\mathrm{mm}$~\cite{Pimenova-Goldobin-EPJE-2014,Pimenova-Goldobin-2016}, the reference pressure increase $\sim10^2\mathrm{Pa}$, which is small compared to the atmospheric pressure, and the integrand of the latter integral remains nearly constant; therefore, $A_{\delta V}\approx P\delta V$. For a quick process of the bubble popping, one can neglect the energy dissipation due to molecular heat conduction and consider the expansion process to be adiabatic; $PV^{c_P/c_V}=const$, where $c_V$ and $c_P$ are the isochoric and isobaric specific vapour heat capacities, respectively, and thus
 $P^{-1}(-\delta P)+(c_P/c_V)V^{-1}\delta V=0$
(here we take into account, that in our terms the pressure change is $(-\delta P)$).
Hence, $A_{\delta V}=(c_V/c_P)V\delta P$.

While the surface tension forces act immediately on the liquid and one can assume the surface tension energy to be released into the kinetic flow energy of the liquid, the expanding vapour enforces both the motion of the surrounding liquid and gas flow. Thus, for the vapour expansion work, one should introduce a coefficient $\vartheta$ of the efficiency of the energy conversion into the mechanical energy of the liquid flow. With this conversion coefficient,
\[
E_\mathrm{bub}=\rho g V_\mathrm{bub}\frac{h}{2}+4\pi r_\mathrm{bub}\sigma+\vartheta A_{\delta V}
=\rho g V_\mathrm{bub}\frac{h}{2}
 +\frac{3\sigma}{r_\mathrm{bub}}V_\mathrm{bub}\vartheta_\Sigma\,.
\]
where we introduce notation
\begin{equation}
\vartheta_\Sigma=1+\frac23\vartheta\frac{c_V}{c_P}.
\label{eq21-01}
\end{equation}

Let us compare the gravitational and surface tension terms in $E_\mathrm{bub}$. They are equal for
\[
h_\ast=\frac{6\sigma\vartheta_\Sigma}{\rho g r_\mathrm{bub}}\,.
\]
For the $n$-heptane--water system, the characteristic observed value of $r_\mathrm{bub}$ was $0.5$--$1\,\mathrm{mm}$~\cite{Pimenova-Goldobin-EPJE-2014,Pimenova-Goldobin-2016}, which yields $h_\ast\sim5$--$10\,\mathrm{cm}$. The case of $h\gg h_\ast$ has been considered in~\cite{Goldobin-Pimenova-2017} and in this paper we assume $h\ll h_\ast$. Henceforth in this paper,
\begin{equation}
E_\mathrm{bub}=\frac{3\sigma\vartheta_\Sigma}{r_\mathrm{bub}}V_\mathrm{bub}\,.
\label{eq21-02}
\end{equation}

\begin{figure}[!t]
\center{
\includegraphics[width=0.6\textwidth]%
{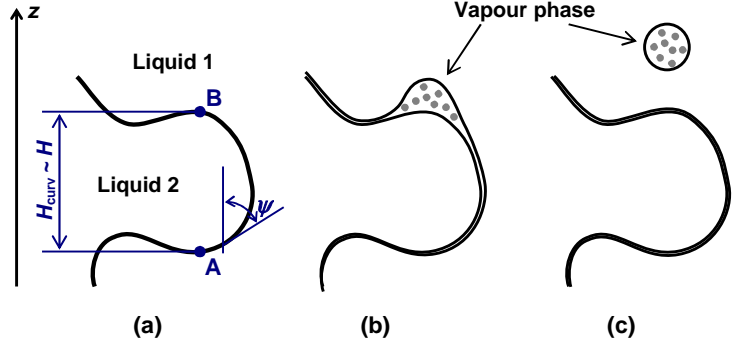}
}
\caption{Process of bubble formation from vapour layer in a well-stirred system
}
\label{fig2}
\end{figure}

Now one has to evaluate $r_\mathrm{bub}$. In~\cite{Pimenova-Goldobin-JETP-2014} for a well-stirred system, the vapour layer thickness reached before the vapour breakaway, which is sketched in figure~\ref{fig2}, has been evaluated
\[
L_\ast=\left(\frac{48\eta_\mathrm{v}}{\rho g}\right)^{1/3}
 S_V^{-2/3}\bigg(\frac{\dot{V}_\mathrm{v}}{V}\bigg)^{1/3},
\]
where $\eta_\mathrm{v}$ is the dynamic viscosity of the vapour.
This vapour breakaway occurs over the interface area $S_\mathrm{bub}$ between points ${\bf A}$ and ${\bf B}$ in figure~\ref{fig2}a, which has the second horizontal scale (perpendicular to the sketch plane) $\sim H$. Thus, $S_\mathrm{bub}\sim H\times(H/\langle\cos\psi\rangle)$, where the averaging of $\cos\psi$ is performed along the interface and, therefore, $\langle\cos\psi\rangle\approx 1/2$. The vapour volume forming a bubble is
\[
V_\mathrm{bub}=S_\mathrm{bub}L_\ast
=2S_V^{-2}L_\ast
 =4\left(\frac{6\eta_\mathrm{v}}{\rho g}\right)^{1/3}
 S_V^{-8/3}\bigg(\frac{\dot{V}_\mathrm{v}}{V}\bigg)^{1/3},
\]
and equation~(\ref{eq21-02}) can be finally rewritten as
\begin{equation}
\frac{E_\mathrm{bub}}{V_\mathrm{bub}}=3\sigma\vartheta_\Sigma
 \left(\frac{\pi}{3}\right)^{1/3}
 \left(\frac{\rho g}{6\eta_\mathrm{v}}\right)^{1/9}
 S_V^{8/9}\bigg(\frac{\dot{V}_\mathrm{v}}{V}\bigg)^{-1/9}.
\label{eq21-03}
\end{equation}

\subsection{Viscous dissipation and heat transfer within the turbulent boundary layer near the interface}
In~\cite{Goldobin-Pimenova-2017}, the flow in the system has been shown to be essentially turbulent. With this turbulent flow momentum and heat transfer in the system is governed by the laws derived for the turbulent boundary layer~\cite{Karman-1930,Prandtl-1933}. According to these laws, as derived in~\cite{Goldobin-Pimenova-2017}, the viscous energy loss rate
\begin{equation}
\frac{\delta\dot{W}_\mathrm{l,k}}{\delta V}\approx
S_V\frac{2\rho u_\ast^3}{\varkappa}\ln\frac{Hu_\ast}{2\xi_0\nu}
\label{eq22-01}
\end{equation}
and the space-average kinetic energy of liquid flow
\begin{equation}
\frac{\delta W_\mathrm{l,k}}{\delta V}
\approx
\frac{\rho u_\ast^2}{2\varkappa^2}\ln^2\frac{Hu_\ast}{2e\xi_0\nu}\,,
\label{eq22-02}
\end{equation}
where empiric constants
\[
\varkappa\approx0.4
\qquad\mbox{ and }\qquad
\xi_0\approx0.13
\]
are known from experiments with turbulent boundary layers, $\nu$ is the characteristic kinematic viscosity of two liquids, $e$ is the Euler's number, and $u_\ast$ measures the characteristic turbulent pulsations of the flow velocity. Physically, $u_\ast$ is determined by flow kinetic energy $W_\mathrm{l,k}$ (equation~(\ref{eq22-02}) provides the relation between them) and governs viscous energy dissipation $\dot{W}_\mathrm{l,k}$ (see equation~(\ref{eq22-01})).

The the heat flux to the interface from one liquid phase can be determined from the temperature boundary layer theory~\cite{Landau-Lifshitz,Schlichting-Gersten} (see~\cite{Goldobin-Pimenova-2017} for details on the implementation of the theory to our case);
\begin{equation}
q_T\approx\frac{\displaystyle\varkappa\rho c_P u_\ast \langle\Theta\rangle}
 {\displaystyle\beta\ln\frac{Hu_\ast}{2\nu}}\,,
\label{eq22-03}
\end{equation}
where turbulent Prandtl number $\beta\approx0.9$ for the turbulent boundary layer is known from experiments, $c_P$ is the liquid specific heat capacity and $\langle\Theta\rangle$ is the average temperature excess above the interfacial boiling point.

\subsection{Approximation of mechanical energy equipartition}
We have to establish the relationship between the flow kinetic energy and the mechanical potential energy in the system. Rising vapour bubbles pump the mechanical energy into the system, while its stochastic dynamics is governed by interplay of its flow momentum and the forces of the gravity and the surface tension on the interface. In thermodynamic equilibrium, the total energy is strictly equally distributed between potential and kinetic energies related to quadratic terms in Hamiltonian (this statement is frequently simplified to a less accurate statement, that energy is equally distributed between kinetic and potential energies associated with each degree of freedom). Being not exactly in the case where one can rigorously speak of thermalization of the stochastic Hamiltonian system dynamics, we still may estimate the kinetic energy of flow to be of the same order of magnitude as the mechanical potential energy of the system. Thus,
\begin{align}
W_\mathrm{l,k}\sim W_\mathrm{l,pg}+W_\mathrm{l,p\sigma}\,,
\nonumber 
\end{align}
where $W_\mathrm{l,pg}$ and $W_\mathrm{l,p\sigma}$ are the gravitational potential energy and the surface tension energy, respectively. In~\cite{Pimenova-Goldobin-EPJE-2014} it has been shown, that for physically plausible regimes of interfacial boiling of a well-stirred system, the surface tension energy is large compared to the gravitational potential energy and the latter can be neglected (as in~\cite{Goldobin-Pimenova-2017});
\begin{align}
W_\mathrm{l,k}\sim W_\mathrm{l,p\sigma}\,.
\label{eq23-01}
\end{align}
The surface tension energy is
\[
W_\mathrm{l,p\sigma}\sim(\sigma_1+\sigma_2)VS_V\,,
\]
where we neglected the interface area of the stratified state compared to the area $VS_V$ in the well-stirred state. Due to the presence of the vapour layer between liquids the effective surface tension coefficient of the interface is $(\sigma_1+\sigma_2)$ but not $\sigma_{12}$ as it would be in the absence of the vapour layer.

\section{Specific interface area $S_V$ providing energy transfer balance}
\subsection{$S_V$ as a function of average overheat $\langle\Theta\rangle$}
From equations~(\ref{eq2-01}), (\ref{eq2-02}), (\ref{eq21-03}) and (\ref{eq22-01}), one finds for $\dot{Q}_V$:
\begin{equation}
\dot{Q}_V\approx(\Lambda_1n_{1}^{(0)}+\Lambda_2n_{2}^{(0)})S_V^{1/8}
 \frac{(3/\pi)^{3/8}}{(3\vartheta_\Sigma)^{9/8}}
 \left(\frac{6\eta_\mathrm{v}}{\rho g}\right)^{1/8}
 \left(\frac{2\rho u_\ast^3}{\varkappa}\ln\frac{u_\ast}{2\xi_0\nu S_V}\right)^{9/8}.
\label{eq31-01}
\end{equation}
The relation between $S_V$ and $u_\ast$ follows from equations~(\ref{eq23-01}) and (\ref{eq22-02});
\[
2\sigma S_V\approx
\frac{\rho u_\ast^2}{2\varkappa^2}\ln^2\frac{u_\ast}{2e\xi_0\nu S_{\!V}}\,,
\]
which can be recast as
\begin{equation}
2\varkappa\sqrt{\sigma/\rho}\,S_V^{1/2}\approx
u_\ast\ln\frac{u_\ast}{2e\xi_0\nu S_{\!V}}\,,
\label{eq31-02}
\end{equation}
On the other hand, the heat inflow $\dot{Q}_V=2q_TS_V$ for a turbulent boundary layer is given by equation~(\ref{eq22-03});
\begin{equation}
\dot{Q}_V
\approx S_V\frac{\displaystyle 2\varkappa\rho c_P u_\ast\langle\Theta\rangle}
 {\displaystyle\beta\ln\frac{u_\ast}{2\nu S_V}}\,.
\label{eq31-03}
\end{equation}
Equations (\ref{eq31-01})--(\ref{eq31-03}) form a self-content equation system for $S_V$, $u_\ast$ and $\dot{Q}_V$. In these equations one can neglect discrepancies in the logarithm arguments as the argument is large $u_\ast/(\nu S_V)\gg1$.

Excluding $\dot{Q}_V$ and logarithms from equations (\ref{eq31-01})--(\ref{eq31-03}), one can find
\[
u_\ast=\left(\frac34\vartheta_\Sigma\right)^\frac92
\left(\frac{\pi}{3}\right)^\frac32\left(\frac{\rho g}{6\eta_\mathrm{v}}\right)^\frac12
\left(\frac{\rho^2c_P}{\beta(\Lambda_1n_{1}^{(0)}+\Lambda_2n_{2}^{(0)})}\right)^4
(\rho\sigma)^{-\frac{17}{4}} \langle\Theta\rangle^4 S_V^{-\frac34}\,.
\]
Substitution of $u_\ast$ into equation~(\ref{eq31-02}) yields
\begin{equation}
S_V=\alpha_{\Theta 1}^{4/5}\langle\Theta\rangle^{16/5}
 \ln^{4/5}\frac{\alpha_{\Theta 2}\langle\Theta\rangle^4}{S_V^{7/4}}\,,
\label{eq31-04}
\end{equation}
where
\begin{gather}
\alpha_{\Theta 1}\equiv\frac{1}{2\varkappa}\left(\frac34\vartheta_\Sigma\right)^\frac92
\left(\frac{\pi}{3}\right)^\frac32\left(\frac{\rho^2 g}{6\eta_\mathrm{v}\sigma}\right)^\frac12
\left(\frac{\rho^2c_P}{\beta(\Lambda_1n_{1}^{(0)}+\Lambda_2n_{2}^{(0)})}\right)^4
(\rho\sigma)^{-\frac{17}{4}}\,,
\label{eq31-05}
\\
\alpha_{\Theta 2}\equiv\frac{1}{2e\xi_0\nu}\left(\frac34\vartheta_\Sigma\right)^\frac92
\left(\frac{\pi}{3}\right)^\frac32\left(\frac{\rho g}{6\eta_\mathrm{v}}\right)^\frac12
\left(\frac{\rho^2c_P}{\beta(\Lambda_1n_{1}^{(0)}+\Lambda_2n_{2}^{(0)})}\right)^4
(\rho\sigma)^{-\frac{17}{4}}\,.
\label{eq31-06}
\end{gather}
One can express the solution to transcendent equation~(\ref{eq31-04}) in the form of a continuous logarithm-fraction~\cite{Goldobin-Pimenova-2017};
\[
S_V=\alpha_{\Theta 1}^{4/5}\langle\Theta\rangle^{16/5}
 \ln^{4/5}
 \frac{\alpha_\Theta\langle\Theta\rangle^{-8/5}}{\displaystyle\ln^{7/5}
 \frac{\alpha_\Theta\langle\Theta\rangle^{-8/5}}{\displaystyle\ln^{7/5}
 \frac{\alpha_\Theta\langle\Theta\rangle^{-8/5}}{\displaystyle
 \dots
 }}}\,,
\]
where $\alpha_\Theta\equiv\alpha_{\Theta 2}/\alpha_{\Theta 1}^{7/5}$.
This result can be rewritten as
\begin{equation}
S_V=\alpha_{\Theta 1}^\frac45\langle\Theta\rangle^\frac{16}{5}
F_{\frac75,\infty}\left(\alpha_\Theta\langle\Theta\rangle^{-\frac{8}{5}}\right)\,,
\qquad
\label{eq31-07}
\end{equation}
where
\[
F_{\frac75,n}(Z)
 \equiv\ln^\frac45\Big( \underbrace{Z\ln^{-\frac75}\Big( \dots Z\ln^{-\frac75}\Big(}_{\mbox{\footnotesize ($n-1$) times}}
 Z\, \Big) \dots \Big)\Big)
\]
for $n=1,2,3,...$\,. Due to a large value of $Z$ the convergence of this function with respect to $n$ is fast; for physically relevant values of $Z$, function $F_{7/5,2}(Z)$ deviates from $F_{7/5,\infty}(Z)$ by less that $1\%$.

\subsection{Systems driven by heat inflow $\dot{Q}_V$}
If the system state is controlled by a given heat inflow $\dot{Q}_V$ (or, equivalently, vapour production rate $\dot{V}_\mathrm{v}/V$, which is related to $\dot{Q}_V$ via equation~(\ref{eq2-01})), one can express from equations (\ref{eq31-01}), (\ref{eq31-02}):
\[
u_\ast=\left(\frac{3\vartheta_\Sigma}{4(\Lambda_1n_{1}^{(0)}+\Lambda_2n_{2}^{(0)})}\right)^\frac12
\left(\frac{\pi}{3}\right)^\frac16\left(\frac{\rho g}{6\eta_\mathrm{v}}\right)^\frac{1}{18}
(\rho\sigma)^{-\frac{1}{4}} \dot{Q}_V^\frac49 S_V^{-\frac{11}{36}}\,.
\]
Then equation~(\ref{eq31-02}) can be rewritten as
\begin{equation}
S_V=\alpha_{Q1}^\frac{36}{29}\dot{Q}_V^\frac{16}{29}
\ln^\frac{36}{29}\left(\alpha_{Q2}\dot{Q}_V^\frac49S_V^{-\frac{47}{36}}\right)\,,
\label{eq32-01}
\end{equation}
where
\begin{gather}
\alpha_{Q1}\equiv\frac{\sqrt{\rho/\sigma}}{2\varkappa}\left(\frac{3\vartheta_\Sigma}{4(\Lambda_1n_{1}^{(0)}+\Lambda_2n_{2}^{(0)})}\right)^\frac12
\left(\frac{\pi}{3}\right)^\frac16\left(\frac{\rho g}{6\eta_\mathrm{v}}\right)^\frac{1}{18}
(\rho\sigma)^{-\frac{1}{4}}\,,
\label{eq32-02}
\\
\alpha_{Q2}\equiv\frac{1}{2e\xi_0\nu}
\left(\frac{3\vartheta_\Sigma}{4(\Lambda_1n_{1}^{(0)}+\Lambda_2n_{2}^{(0)})}\right)^\frac12
\left(\frac{\pi}{3}\right)^\frac16\left(\frac{\rho g}{6\eta_\mathrm{v}}\right)^\frac{1}{18}
(\rho\sigma)^{-\frac{1}{4}}\,.
\label{eq32-03}
\end{gather}

Solution to equation~(\ref{eq32-01}) can be found in the form of a continuous logarithm-fraction;
\begin{equation}
S_V=\alpha_{Q1}^\frac{36}{29}\dot{Q}_V^\frac{16}{29}
F_{\frac{47}{29},\infty}\left(\alpha_Q\dot{Q}_V^{-\frac{8}{29}}\right)\,,
\qquad
\label{eq32-04}
\end{equation}
where $\alpha_Q\equiv\alpha_{Q2}/\alpha_{Q1}^{47/29}$ and
\[
F_{\frac{47}{29},n}(Z)
 \equiv\ln^\frac{36}{29}\Big( \underbrace{Z\ln^{-\frac{47}{29}}\Big( \dots Z\ln^{-\frac{47}{29}}\Big(}_{\mbox{\footnotesize ($n-1$) times}}
 Z\, \Big) \dots \Big)\Big)
\]
for $n=1,2,3,...$\,.

\section{Conclusion}
On the basis of the energy flux balance in the system, the assumption of the system stochastization and transport laws for turbulent boundary layers, we have evaluated the specific interface area $S_V$ as a function of parameters which can control the system state: average overheat $\langle\Theta\rangle$ and heat inflow $\dot{Q}_V$ (or, equivalently, vapour generation rate $\dot{V}_\mathrm{v}/V$). The results are derived in the form of continuous logarithm-fractions~(\ref{eq31-07}) and (\ref{eq32-04}), which possess fast convergence properties (with respect to the number of iterations) for physically relevant value of parameters~\cite{Goldobin-Pimenova-2017}.

\ack{
The work has been financially supported by the Russian Foundation for Basic Research (grant no.\ 15-01-04842).}


\section*{References}
\begin{thebibliography}{20}
\bibitem{Krell-1982}
 Krell E 1982
 {\em Handbook of Laboratory Distillation}
 (Amsterdam: Elsevier) chapter 4.3

\bibitem{Geankoplis-2003}
 Geankoplis C J 2003
 {\em Transport Processes and Separation Process Principles}
 (New Jersey: Prentice Hall)

\bibitem{Simpson-etal-1974}
 Simpson H C, Beggs G C and Nazir M  1974
 {\em Desalination} {\bf 15} 11--23

\bibitem{Celata-1995}
 Celata G P, Cumo M, D'Annibale F, Gugliermetti F and Ingui' G 1995
 {\em Int.\ J.\ Heat Mass Transfer} {\bf 38} 1495--504

\bibitem{Roesle-Kulacki-2012-1}
 Roesle M L and Kulacki F A
 {\em Int.\ J.\ Heat Mass Transfer} {\bf 55} 2160--5

\bibitem{Roesle-Kulacki-2012-2}
 Roesle M L and Kulacki F A
 {\em Int.\ J.\ Heat Mass Transfer} {\bf 55} 2166--72

\bibitem{Sideman-Isenberg-1967}
 Sideman S and Isenberg J 1967
 {\em Desalination} {\bf 2} 207--14

\bibitem{Kendoush-2004}
 Kendoush A A 2004
 {\em Desalination} {\bf 169} 33--41


\bibitem{Filipczak-etal-2011}
 Filipczak G, Troniewski L and Witczak S 2011
 {\em Evaporation, Condensation and Heat transfer}
 ed Ahsan A (Rijeka: InTech) p 123--50

\bibitem{Gordon-etal-1961}
 Gordon K F, Singh T and Weissman E Y 1961
 {\em Int.\ J.\ Heat and Mass Transfer} {\bf 3} 90--3

\bibitem{Prakash-Pinder-1967-1}
 Prakash C B and Pinder K L 1967
 {\em Can.\ J.\ Chem.\ Engineering} {\bf 45} 210--4

\bibitem{Prakash-Pinder-1967-2}
 Prakash C B and Pinder K L 1967
 {\em Can.\ J.\ Chem.\ Engineering} {\bf 45} 215--20
	
	
\bibitem{Pimenova-Goldobin-JETP-2014}
 Pimenova A V and Goldobin D S 2014
 {\em JETP} {\bf 119} 91--100
	
\bibitem{Pimenova-Goldobin-EPJE-2014}
 Pimenova A V and Goldobin D S 2014
 {\em Eur.\ Phys.\ J.} E {\bf 37} 108

\bibitem{Goldobin-Pimenova-2017}
 Goldobin D S and Pimenova A V 2017
 Specific interface area and self-stirring in a two-liquid system experiencing intense interfacial boiling below the bulk boiling temperatures of both components
 {\em Eur.\ Phys.\ J.} ST {\it unpublished} ({\it Eprint} arXiv:1606.09520)

\bibitem{Pimenova-Goldobin-2016}
 Pimenova A V and Goldobin D S 2016
 {\em J.\ Appl.\ Mech.\ Tech.\ Phys.} {\bf 57} 1182--9


\bibitem{Karman-1930}
 von Karman Th 1930
 Mechanische \"Ahnlichkeit und Turbulenz
 {\em Nachrichten von der Gesellschaft der Wissenschaften zu G\"ottingen, Fachgruppe 1 (Mathematik)}
 {\bf 5} 58--76
 ({\it English translation}: von Karman Th 1931 Mechanical Similitude and Turbulence
 {\em Tech.\ Mem.\ NACA} {\bf 611})

\bibitem{Prandtl-1933}
 Prandtl L 1933
 Neuere Ergebnisse der Turbulenzforschung
 {\em Z.\ Ver.\ dtsch.\ Ing.} {\bf 77} 105--14
 ({\it English translation}: Prandtl L 1933 Recent results of turbulence research
 {\em Tech.\ Mem.\ NACA} {\bf 720})

\bibitem{Landau-Lifshitz}
 Landau L D and Lifshitz E M 1986
 {\em Fluid Mechanics}
 (Moscow: Nauka)

\bibitem{Schlichting-Gersten}
 Schlichting H and Gersten K 2000
 {\em Boundary-Layer Theory}
 (Springer)

	
\end{thebibliography}
\end{document}